\documentclass[conference]{IEEEtran}

\usepackage{subfigure}
\usepackage[dvipsnames]{xcolor}
\usepackage{amsmath}
\usepackage{amssymb}
\usepackage{amsthm}
\usepackage{comment}
\usepackage{color}
\usepackage{mathtools}
\usepackage{algorithm,algorithmic}
\usepackage{bbold}
\usepackage{datetime}
\usepackage{graphicx}
\usepackage{cuted}
\usepackage{tikz,pgfplots,pgfplotstable,pgf-pie}
\usepackage{xfrac}
\usepackage{mwe}
\usetikzlibrary{patterns}
\usepackage{verbatim}
\usepackage[dvipsnames]{xcolor}
\usepackage{epstopdf}
\usepackage{eqparbox}

\newcommand{\cw}{CW\textsubscript{min}}
\newcommand{\CW}{CW\textsubscript{min} }

\begin{document}
\IEEEoverridecommandlockouts
\IEEEpubid{\begin{minipage}{\textwidth}\ \\[10pt]
\centering{\copyright2021 IEEE. Personal use of this material is permitted.  Permission from IEEE must be obtained for all other uses, in any current or future media, including reprinting/republishing this material for advertising or promotional purposes, creating new collective works, for resale or redistribution to servers or lists, or reuse of any copyrighted component of this work in other works.}
\end{minipage}}

\title{\CW Estimation and Collision Identification in Wi-Fi Systems}

\author{\IEEEauthorblockN{Amir-Hossein Yazdani-Abyaneh and Marwan Krunz}

\IEEEauthorblockA{Department of Electrical and Computer Engineering, University of Arizona, AZ, USA\\
Email: \{yazdaniabyaneh, krunz\}@email.arizona.edu} 
}

\maketitle


\begin{abstract}
Wi-Fi networks are susceptible to aggressive behavior caused by selfish or malicious devices that reduce their minimum contention window size (\cw) to below the standard \cw . In this paper, we propose a scheme called \emph{Minimum Contention Window Estimation} (CWE) to detect aggressive stations with low \cw 's, where the AP estimates the CW\textsubscript{min} value of all stations transmitting uplink by monitoring their backoff values over a period of time and keeping track of the idle time each station spends during backoff. To correctly estimate each backoff value, we present a cross-correlation based technique that uses the frequency offset between the AP and each station to identify stations involved in uplink collisions. The AP constructs empirical distributions for the monitored backoff values and compares them with a set of nominal PMF’s, created via Markov analysis of the DCF protocol to estimate CW\textsubscript{min} of various stations. After detecting the aggressive stations, the AP can choose to stop serving those stations. Simulation results show that the accuracy of our collision detection technique is $96\%$, $94\%$, and $88\%$ when there are 3, 6, and 9 stations in the WLAN, respectively. For the former WLAN settings, the estimation accuracy of CWE scheme is $100\%$, $98.81\%$, and $96.3\%$, respectively.
\end{abstract}

\section{Introduction}\label{sc:Introduction}
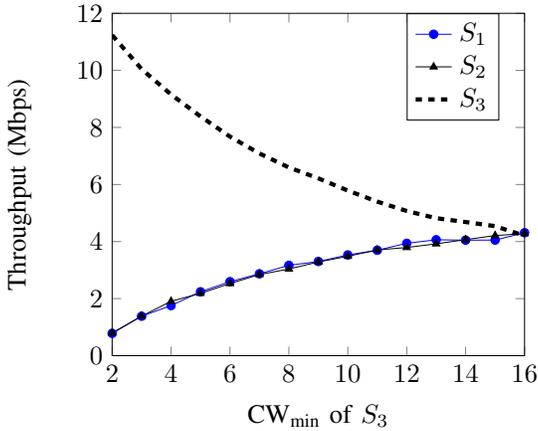
\begin{figure}[h]

\begin{tikzpicture}[scale=0.8, transform shape]
\begin{axis}[title={},xlabel={\CW of $S_3$},
    ylabel={Throughput (Mbps)},
    xmin=2, xmax=16,
    ymin=0, ymax=12,
    legend style={at={(axis cs:12,12)},anchor=north west}
    ,nodes={scale=1.25, transform shape}]

\addplot[color=blue,mark=*]coordinates{(2,0.7820551200000001)(3,1.3834374400000002)(4,1.7551072)(5,2.24034272)(6,2.5939452)(7,2.87011648)(8,3.16693608)(9,3.30115016)(10,3.5308627200000005)(11,3.6960492799999995)(12,3.93866704)(13,4.06255696)(14,4.049651760000001)(15,4.0522328000000005)(16,4.3103368)
};
\addlegendentry{$S_1$ }

\addplot[color=black,mark=triangle*]coordinates{(2,0.79754136)(3,1.3860184800000002)(4,1.9048075199999999)(5,2.18614088)(6,2.5294192000000004)(7,2.84430608)(8,3.03788408)(9,3.2856639199999997)(10,3.484404)(11,3.70637344)(12,3.7915477600000003)(13,3.9205997600000004)(14,4.06771904)(15,4.20967624)(16,4.27936432)

};
\addlegendentry{$S_2$ }

 \addplot [color=black, dashed, line width=2pt] coordinates {(2,11.23526712)(3,10.04540768)(4,9.154948880000001)(5,8.385798959999999)(6,7.6785939999999995)(7,7.087535839999999)(8,6.597138240000001)(9,6.215144319999999)(10,5.78669168)(11,5.3995356800000005)(12,5.0717436)(13,4.81880168)(14,4.6845875999999995)(15,4.540049359999999)(16,4.2045141600000004)
};
 \addlegendentry{$S_3$ };

\end{axis}
\end{tikzpicture}

  \caption{Per-station throughput for a network of three stations vs. \CW of $S_3$ (\CW$=16$ for $S_1$ and $S_2$). } 
    \label{fig:fp}
  
  \end{figure}

Wi-Fi end-users, also known as stations, demand fair allocation of the channel airtime.  The 802.11 MAC  protocol \cite{80211}, known as the Distributed Coordination Function (DCF), uses Carrier Sense Multiple Access with Collision Avoidance (CSMA/CA) with exponential backoff to provide fair channel access in a distributed manner, provided that all stations comply with the DCF protocol. Under DCF, a station that wants to transmit must first  sense the channel for a fixed duration, called the DCF initial Inter-Frame Space (DIFS). If the channel is sensed to be idle during the DIFS period, the station starts its transmission; otherwise, the station defers its transmission and waits for a random backoff period. The backoff period consists of $k$ idle slots, where $k$ is randomly chosen from $\{0,1,...,\text{CW}-1\}$. Initially, CW is set to a default value CW\textsubscript{min}. After a collision, \CW is doubled until it reaches the maximum allowable contention window (CW\textsubscript{max}). Generally, a station that has consecutively collided for $j$ times chooses its $k$ randomly from $\{0,1,...,\text{min}(2^{j}\text{CW\textsubscript{min}}, \text{CW\textsubscript{max}})-1\}$. The exponential increase in CW helps stations avoid collisions. Following a successful transmission, a station resets its CW to CW\textsubscript{min}.\\
 In a Wi-Fi system, aggressive behavior for channel access can be attributed to malicious reasons  to degrade the network's performance \cite{Xu05,toledo08,manzo05,raymond09} or it can be caused by  selfish stations that try to gain more access to channel airtime \cite{Li15,radosavac05,Kyasanur05,lu10}. An example of malicious behavior is channel jamming attacks \cite{Xu05,law09}, which can be considered as a particular type of Denial-of-service (DoS) attack \cite{toledo08, khatib10,gupta02}. In addition to transmitting a high-power signal to disrupt other users' transmissions, a malicious station can also transmit fake packets to prevent normal users from communicating \cite{Li15}. In contrast, a selfish station alters its protocol parameters to get an unfair share of the channel airtime at the expense of other well-behaving stations. For example, this station may reduce the value of its SIFS or DIFS below the standard values. It may choose a larger value of the remaining transmission duration field in the MAC header to force other stations to back off for longer periods. It may also lower its \CW so that it captures the channel more often than other stations. Although DCF does an excellent job in ensuring fairness among devices and reducing collisions, it is still vulnerable to aggressive stations that do not abide by the standard protocol, hence harming the performance of compliant stations.\\
To cast more light on this issue, consider\IEEEpubidadjcol a Wi-Fi network with three backlogged stations, all following the DCF protocol with access category $A_3$ \cite{80211}. All stations are in each other's sensing range. Stations $S_1$ and $S_2$ select a standard \CW of 16. In Figure \ref{fig:fp}, we show the per-station throughput for different values of $S_3$'s \cw . Whenever \CW of $S_3$ is less than 16, $S_1$ and $S_2$ have lower throughputs than $S_3$.  Our goal is to enable the AP to detect aggressive stations by estimating their \CW and comparing them with a standard defined value.
 The problem of detecting stations with low \CW values has been studied in the literature, as described in Section \ref{sc:relatedworks}. However, prior works propose protocol modifications \cite{Kyasanur05,konorski01,konorski02,cagalj04}, assume backlogged stations \cite{Li15,Rong06,Tang14, Raya06, AmirMohammed2019DySPAN}, or they do not consider the hidden terminal problem \cite{Li15, Raya06, AmirMohammed2019DySPAN}. In this paper, we consider the problem of \emph{detecting} aggressive stations with low \CW setting, but without imposing any computational overhead on any station. Our approach is only implemented at the AP, but without altering the DCF protocol. We make no assumptions about the traffic type of any station. Further, we take into consideration the hidden terminal problem by introducing a new correlation-based technique for collision detection.  
Our \emph{Minimum Contention Window Estimation} (CWE) has two phases, a monitoring phase and an estimation phase. In the monitoring phase, the AP monitors transmission activity of each station and notes down the idle durations in which stations decrease their backoff counters. The AP translates the monitored idle durations to their representing backoff values and constructs an empirical distribution (PMF) of these backoff values.  The AP makes sure that the periods of time each station is idle is due to its backoff process and not caused by an empty transmission buffer. We assume that all transmitted frames include the \texttt{Queue Size} subfield of the \texttt{QoS Control} field of the MAC header \cite{80211}\footnote{IEEE 802.11 standard requires all QoS data packets to include the \texttt{Queue Size} subfield in their MAC headers.}. Thus, the AP derives backoff values only for packets that their prior transmission indicated a non-empty transmission  buffer, i.e. non-zero \texttt{Queue Size} value. Another important aspect of correctly estimating the backoff value is for the AP to detect collisions in the uplink and identify stations that are involved in a collision. Most prior works related to misbehavior detection do not consider the possibility of hidden terminals, and if they do, they discard any observation related to a collision.  To determine the identities of colliding stations, we present a correlation-based technique that uses the frequency offset (FO) between each station and the AP. For each station-AP pair, the technique calculates the cross-correlation of a collided signal with the 802.11 preamble that is modified to include the FO effects of the station-AP pair,  and looks for peaks in the cross-correlation that are higher than a given threshold to identify colliding stations.\\
  In the second phase of CWE, the constructed PMF's are compared with a set of nominal PMF's, which are derived based on Markov Chain (MC) analysis of the CSMA/CA protocol \cite{bianchi00} whereby all stations but one are compliant and the \CW of the non-compliant station is changed within a range to construct different nominal PMF's (one per \CW value of the non-compliant station). The observed and nominal PMF's are compared using Jensen-Shannon divergence measure. The \CW with a nominal PMF of least divergence measure with the observed PMF is taken as the estimated \CW for the station under observation. Stations with estimated \CW values lower than the standard value are considered as aggressors.\\
 Simulation results with three, six, and nine stations show that our collision detection technique achieves an accuracy of $96 \%$, $94\%$, and $88\%$, respectively. The corresponding accuracy of the \cw -estimation algorithm is $100\%$, $98.81\%$, and $96.3\%$, respectively. 
The paper is organized as follows. In Section \ref{sc:cwe}, we introduce CWE along with the backoff value estimation algorithm.  Our collision identification technique and evaluation results are presented in Sections \ref{sc:collision_detection} and \ref{sc:evalualtions}, respectively. Finally, we survey related works and conclude the paper in Sections \ref{sc:relatedworks} and \ref{sc:conclusions}, respectively.

\section{Minimum Contention Window Estimation (CWE)}\label{sc:cwe}
Our system model includes a WLAN with $N$ stations and an AP. We denote the $j$th station by $S_j$ and the standard \CW by $W_s$.  To estimate \CW of an arbitrary station, say $S_j$, the AP tracks the backoff values selected by $S_j$ over an observation period $T$. The set of backoff values selected by $S_j$ are denoted by $K_j=[K_j(1),K_j(2),...,K_j(L_j)]$, where $K_j(i)$ is the $i$th backoff value selected by $S_j$ during $T$ and $L_j$ is the total number of selected backoff values by $S_j$. In Section \ref{sc:backoff}, we explain the process of obtaining the vector $K_j$. For now, we assume that $K_j(i)$'s have been estimated by the AP. The AP constructs an empirical probability mass function (PMF) from the vector $K_j$, as:
\begin{equation}
H_j(n)=\frac{\sum_{i=1}^{L_j}\mathbb{1}(K_j(i)==n)}{|K_j|},\quad  n\in \{0,1,...,2^{M}W_s-1\}
\label{eq:hist}
\end{equation}
where $M$ is the maximum number of allowed retransmissions.
Consider an arbitrary station $S \in \{S_1,...,S_N \}$ with \cw$=W$. If $S$ is compliant, then $W$ is the standard value. Different \CW settings for $S$ result in different backoff values, thus different PMF's. After each successful transmission, $S$ samples its backoff values from a uniform distribution $U_{[0,W-1]}$. Because stations double their contention window after a collision, $S$ will select its backoff values from $U_{[0,2W-1]}$ after any collision that follows a successful transmission. Considering the possibility of collisions, the overall PMF of the backoff values selected by $S$, denoted by $H$, depends on $W$ and the collision probability in the WLAN.\\
 The AP maintains a set of nominal PMF's, denoted by $\mathcal{P}^{(N)}=\{P_2^{(N)},P_3^{(N)},...,P_{W_{s}}^{(N)}\}$, where $P^{(N)}_{l}$ is the PMF of selected backoff values of a station with a \cw$=l$ in a WLAN of $N$ stations, where all other $N-1$ stations have a \cw$=W_s$. To obtain $W$, the AP compares $H$ with each $P^{(N)}_{l}$ for $l \in \{2, 3, ..., W_s\}$, the $l$ with the nominal PMF of $P^{(N)}_{l}$ that has the least difference from $H$ is considered as the estimated \CW for station $S$. If $S$ does not collide during its transmissions, then it will always randomly select a backoff value $k$ from the uniform distribution $d_0=U_{[0,W-1]}$. On the other hand, if $S$ is involved in $j$ consecutive collisions, it will randomly select $k$ from $d_j=U_{[0,2^{j}W-1]}$. Therefore, the overall PMF of backoff value selections of a station with \cw$=W$, i.e. $P^{(N)}_{W}$, is a composition of $d_0,d_1,...,d_M$, where $M$ is the maximum number of allowed retransmissions. To derive $P_W^{(N)}$, we define a random variable $X$ that represents the backoff stage of $S$. $X$ takes values from $\{0,1,...,M\}$. For instance, $X=j$ means that $S$ has collided  $j$ consecutive times and will randomly select its backoff value from $\{0,1,...,2^{j}W-1\}$. The overall distribution of  $k$ is the weighted superposition of $d_0, d_1, ..., d_M$: 

\begin{equation}
P^{(N)}_W=\sum_{i=0}^{M}\Pr [X=i]\times d_i.
\label{eq:e5}
\end{equation}
 To find $\Pr [X]$'s, we need to find the collision probabilities for when $S$ selects a \cw$=W$. In \cite{Rong06}, authors obtain collision and packet transmission probabilities for different \CW settings each station in the WLAN. Their analysis is based on  Bianchi's MC modeling of the 802.11's DCF protocol \cite{bianchi00}. For this work we only need to obtain the former probabilities for when all stations except one are compliant (\cw$=W_s$). This simplification does not degrade the performance of CWE; instead, it further reduces the computational complexity of CWE from $O(N^{W_s-1})$ to $O(W_s-1)$.\\
  Bianchi developed a bidimensional MC for a WLAN with $N$ stations, where all stations have a  \cw$= W_s$, and assumed that the collision probability is constant, denoted by $p$. Let $\{s(t), b(t)\}$ represent the state of the MC, where $s(t) \in \{0,1,...,M\}$ is a stochastic process that represents the backoff stage of a station, and $b(t)$ is a stochastic process that represents the backoff counter for the station. At a stage $s(t)=i$, $i \in \{0,1,...,M\}$, $b(t)$ can take values from the set $\{0,1,...,W_i-1\}$, where $W_i=2^{i}W_s$. The one step transition probabilities are represented as:

\begin{equation}
    \begin{cases}
      \Pr[i,k|i,k+1]=1 				& k\in \{0,...,W_i-2\}\quad  i \in \{0,...,M\} \\
      \Pr[0,k|i,0]=\frac{1-p}{W_s}	& k\in \{0,...,W_s-1\}\quad  i \in \{0,...,M\} \\     
      \Pr[i,k|i-1,0]=\frac{p}{W_i}	& k\in \{0,...,W_s-1\}\quad  i \in \{0,...,M\} \\
      \Pr[M,k|M,0]=\frac{p}{W_M}	& k\in \{0,...,W_M-1\},  \\
    \end{cases}       
    \label{eq:e7}
\end{equation}
where $ \Pr[i,k|j,l]=\Pr[s(t+1)=i,b(t+1)=k|s(t)=j,b(t)=l]$. Deriving the steady state probabilities, The probability of a transmission in a randomly chosen time slot, denoted by $\tau$ is:
\begin{equation}
\tau=\frac{2(1-2p)}{(1-2p)(W_s+1)+pW_s(1-(2p)^M)}.
\label{eq:e8}
\end{equation}
And the probability of a collision can be represented as:

\begin{equation}
p=1-(1-\tau)^{N-1}.
\label{eq:e9}
\end{equation}
Equations \ref{eq:e8} and \ref{eq:e9} form two nonlinear equations with two unknowns which can be solved numerically to obtain $p$ and $\tau$. It is important to note that Equations \ref{eq:e8} and \ref{eq:e9} are only valid when all the $N$ stations have $\text{CW}_{\text{min}}=W_s$, which is not the case in our system model, since an aggressor has a lower \CW value than the standard value. To calculate the collision probability for $S$ with \cw $=W$, which is needed to obtain the proper $\Pr [X]$'s in Equation \ref{eq:e5}, we assume that all $N-1$ other stations have a \cw$=W_s$. We denote the transmission and collision probabilities for $S$ by $\tau$ and $p$, respectively. The transmission and collision probabilities for the compliant stations (\cw$=W_s$) are denoted by $\tau ^{'}$ and $p^{'}$, respectively. Following the analysis presented in \cite{Rong06} the probabilities can be obtained by solving the following four nonlinear equations:

\begin{equation}
    \begin{cases}
    \tau=\frac{2(1-2p)}{(1-2p)(W+1)+pW(1-(2p)^M)}	 	\\
 	p=1-(1-\tau^{'})^{N-1}		\\     
	\tau ^{'}=\frac{2(1-2p^{'})}{(1-2p^{'})(W_s+1)+p^{'}W_s(1-(2p^{'})^M)}			\\
 	p^{'}=1-(1-\tau)(1-\tau^{'})^{N-2}.				\\      
    \end{cases} 
   \label{eq:e10}
\end{equation}

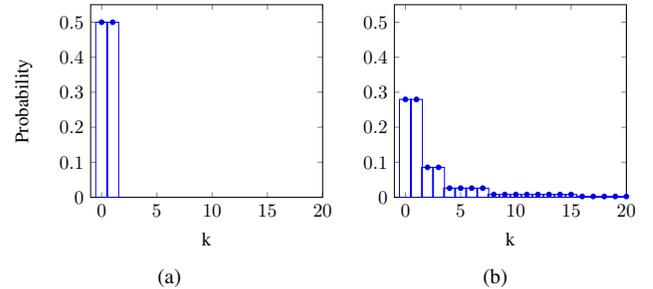
\begin{figure}[h]
  \subfigure[]{
\begin{tikzpicture}[scale=0.45, transform shape]
\begin{axis}[title={},xlabel={k},
    ylabel={Probability},
       xmin=-1, xmax=20,
       ymin=0, ymax=0.55,
nodes={scale=1.6, transform shape}]
\addplot+[ybar] plot coordinates
	{(0,0.5)(1,0.5)

};	
\end{axis}
\end{tikzpicture}
  \label{fig:h1}%
}\subfigure[]{
 \begin{tikzpicture}[scale=0.45, transform shape]
\begin{axis}[title={},xlabel={k},
    ylabel={},
       xmin=-1, xmax=20,
       ymin=0, ymax=0.55,
nodes={scale=1.6, transform shape}]
\addplot+[ybar] plot coordinates
	{(0,0.2795075539369487)(1,0.2795075539369487)(2,0.08551704927087882)(3,0.08551704927087882)(4,0.02615411283844035)(5,0.02615411283844035)(6,0.02615411283844035)(7,0.02615411283844035)(8,0.007988490619209677)(9,0.007988490619209677)(10,0.007988490619209677)(11,0.007988490619209677)(12,0.007988490619209677)(13,0.007988490619209677)(14,0.007988490619209677)(15,0.007988490619209677)(16,0.0024296377314760764)(17,0.0024296377314760764)(18,0.0024296377314760764)(19,0.0024296377314760764)(20,0.0024296377314760764)(21,0.0024296377314760764)(22,0.0024296377314760764)(23,0.0024296377314760764)(24,0.0024296377314760764)(25,0.0024296377314760764)(26,0.0024296377314760764)(27,0.0024296377314760764)(28,0.0024296377314760764)(29,0.0024296377314760764)(30,0.0024296377314760764)(31,0.0024296377314760764)(32,0.0007285759646651563)(33,0.0007285759646651563)(34,0.0007285759646651563)(35,0.0007285759646651563)(36,0.0007285759646651563)(37,0.0007285759646651563)(38,0.0007285759646651563)(39,0.0007285759646651563)(40,0.0007285759646651563)(41,0.0007285759646651563)(42,0.0007285759646651563)(43,0.0007285759646651563)(44,0.0007285759646651563)(45,0.0007285759646651563)(46,0.0007285759646651563)(47,0.0007285759646651563)(48,0.0007285759646651563)(49,0.0007285759646651563)(50,0.0007285759646651563)(51,0.0007285759646651563)(52,0.0007285759646651563)(53,0.0007285759646651563)(54,0.0007285759646651563)(55,0.0007285759646651563)(56,0.0007285759646651563)(57,0.0007285759646651563)(58,0.0007285759646651563)(59,0.0007285759646651563)(60,0.0007285759646651563)(61,0.0007285759646651563)(62,0.0007285759646651563)(63,0.0007285759646651563)(64,0.0002080349118715034)(65,0.0002080349118715034)(66,0.0002080349118715034)(67,0.0002080349118715034)(68,0.0002080349118715034)(69,0.0002080349118715034)(70,0.0002080349118715034)(71,0.0002080349118715034)(72,0.0002080349118715034)(73,0.0002080349118715034)(74,0.0002080349118715034)(75,0.0002080349118715034)(76,0.0002080349118715034)(77,0.0002080349118715034)(78,0.0002080349118715034)(79,0.0002080349118715034)(80,0.0002080349118715034)(81,0.0002080349118715034)(82,0.0002080349118715034)(83,0.0002080349118715034)(84,0.0002080349118715034)(85,0.0002080349118715034)(86,0.0002080349118715034)(87,0.0002080349118715034)(88,0.0002080349118715034)(89,0.0002080349118715034)(90,0.0002080349118715034)(91,0.0002080349118715034)(92,0.0002080349118715034)(93,0.0002080349118715034)(94,0.0002080349118715034)(95,0.0002080349118715034)(96,0.0002080349118715034)(97,0.0002080349118715034)(98,0.0002080349118715034)(99,0.0002080349118715034)(100,0.0002080349118715034)

};	
\end{axis}
\end{tikzpicture}
  \label{fig:h2}

}
  \caption{Backoff value distribution of $S$ with $W=2$, $M=7$, and $N=10$ where $S$ (a) does not  and (b) does double its contention window in the case of collisions.} 
  \label{fig:h1_h2}
 
  \end{figure}

In Equation \ref{eq:e5}, $\Pr[X]$'s are needed to be calculated to construct nominal PMF's ($P^{(N)}_W$'s). $\Pr[X=i]$ can be interpreted as the steady state probability of being in a backoff stage $i$ ($\Pr [s(t)=i]$), and it can be calculated as:

\begin{equation}
\Pr [X=i]=    
    \begin{cases}
    1-p & i=0	 	\\
 	(1-p)p^{i} & i=1,...,M-1\\ 
 	 p^{M} & i=M\\      
    \end{cases} 
    \label{eq:e10_2}
\end{equation}
In Figures \ref{fig:h1} and \ref{fig:h2}, we show the PMF of backoff value selections of $S$ when $W=2$, $N=10$, and $M=7$, where $S$ does not and does double its contention window in case of collisions, respectively. To obtain Figure \ref{fig:h2}, we solve Equation \ref{eq:e10} and calculate $p=0.612$.
To compare the constructed empirical PMF (i.e., $H$) with each $P^{(N)}_l\in \mathcal{P}^{(N)}$ for $l\in \{2,3,...,W_s\}$, we use Jensen-Shannon divergence \cite{cha08}, which is based on Shannon's concept of uncertainty (entropy), to measure the similarity between two probability distributions. The Jensen-Shannon divergence measure between two PMF's $H$ and $P$ is denoted by $J(H,P)$, and it is calculated as:
\begin{multline*}
J(H,P)=\frac{1}{2} \biggl[ \sum_{i=1}^{|H|}P(i)ln\left(\frac{2P(i)}{P(i)+H(i)}\right)\\ +
 \sum_{i=1}^{|H|}H(i)ln\left(\frac{2H(i)}{P(i)+H(i)}\right)\biggl],
    \label{eq:e11}
\end{multline*}

where $P(i)$ and $H(i)$ are the $i$th elements of $P$ and $H$, respectively, and $|.|$ is the cardinality operator. The estimated \CW value for $S_j$, i.e. $W_j$, can be estimated as:
 \begin{equation}
 W_{j}=\operatorname*{argmin}_{l \in \{2,3,...,W_s\}} J(H_j,P^{(N)}_l).
    \label{eq:e12}
\end{equation}
\subsection{Backoff Counter Estimation}\label{sc:backoff}
\begin{figure}
 \centering
\includegraphics[scale=0.65]{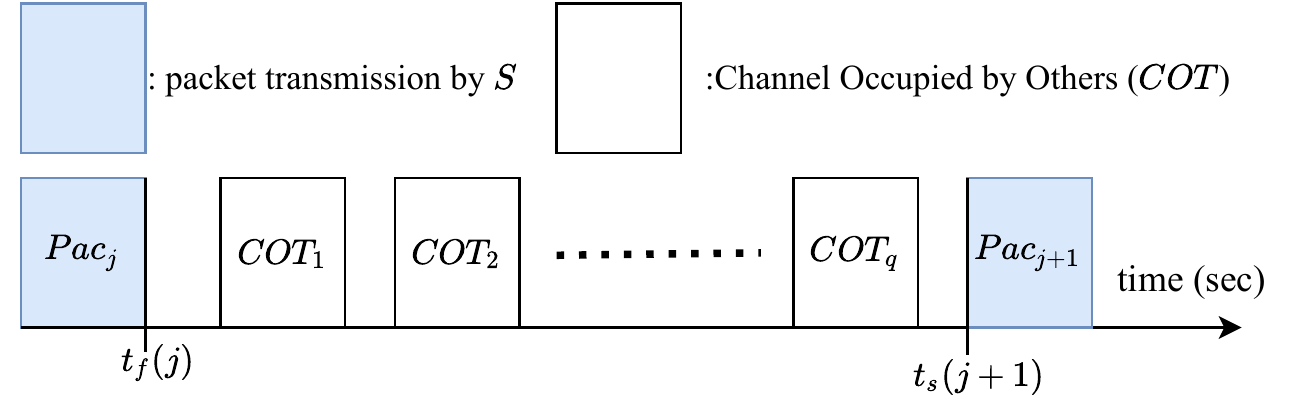}
\caption{An example of gathering observations for estimating the $(j+1)$th backoff value selected by $S$.}
\label{fig:backoff}
\end{figure}

To estimate backoff values selected by $S$ for each channel access attempt, the AP monitors $S$'s transmission activity and notes down the idle durations in which $S$ decreased its backoff counter. Afterwards, the monitored idle durations are translated to backoff values which caused those specific idle periods. During the monitoring period, the AP needs to be accurate in sensing $S$'s transmission, hence it needs to detect any uplink collisions and identify stations involved in them.  We tackle the former by introducing \textit{Collision Identification Technique} (CIT) that helps identify all stations involved in an uplink collision. Using CIT, the AP will be able to monitor the channel and associate each channel busy time to a subset of stations. We explain CIT in Section \ref{sc:collision_detection}. Also, the AP needs to make sure that the duration $S$ spent in an idle states was due to decreasing its backoff counter and not caused by an empty transmission buffer. The former is always true when the WLAN is in a saturated traffic scenario, where stations are backlogged with packets to transmit.  To tackle the stated challenge, the AP will use the information that QoS packets must include in their QoS Data Field of their MAC headers, namely the \texttt{Queue Size} subfield \cite{80211}. The \texttt{Queue Size} subfield indicates the number of bytes that are present in the queue of the transmitter at the time of transmission. Therefore, a none-zero \texttt{Queue Size} value will suggest that the transmitter entered backoff stage immediately after that packet transmission, hence all sensed idle durations were due to decreasing the backoff counter. Also, we know that a packet that experiences collision will be set for retransmission for at most an $M$ number of retransmissions, this means that the  \texttt{Queue Size} of packets that are inside a collision will be considered as nonzero for backoff value estimation, too. Nonetheless, there is a low possibility that a packet might fail to successfully transmit for $m$ times and get dropped. In this case, if there are no packets left to transmit at the buffer of $S$, on average, our algorithm will mistakenly measure the backoff value to be more than $\frac{2^{m+1}-1}{2}W_s$. For this case, the AP will disregard that backoff value estimation.\\
We explain how each backoff value is estimated during two successive packet transmissions of $S$. We denote the $j$th packet transmission by $S$ during $T$ by $Pac_j$. Also, we  define $COT_{j+1}(i)$ to be the $i$th duration of time that the channel becomes occupied by stations other than $S$ during the contention period for transmitting the $(j+1)$th packet of $S$. Figure \ref{fig:backoff} shows an example of observations seen during two packet transmissions, i.e. $Pac_j$ and $Pac_{j+1}$. To find the value of $K(j+1)$ (i.e., the $(j+1)$th backoff value selected by $S$ during $T$), the AP has to mark the time instant of the end of $Pac_j$'s transmission, denoted by $t_f(j)$, and the time instant $Pac_{j+1}$ started getting transmitted, denoted by $t_s(j+1)$. We assume that $Pac_j$ has a non-zero value for its \texttt{Queue Size} subfield; otherwise, the AP would have disregarded the estimation of $K(j+1)$. It is important to note that either $Pac_j$ or $Pac_{j+1}$ may be involved in collisions; however, using CIT the AP will be able to determine the start of packet transmissions by different stations in a collision, and by assuming fixed packet sizes, the AP can estimate the end of a packet transmission in a collision, too. Therefore, our backoff estimation example holds for the case of collisions, too. The value of $K(j+1)$ can be calculated as:
 \begin{equation}
 K(j+1)=\frac{t_s(j+1)-t_f(j)-\sum_{i=1}^{q}COT_i-q\times T_{DIFS}}{T_{MAC}},
    \label{eq:backoff}
\end{equation}
where $T_{MAC}$ and $T_{DIFS}$ are the MAC time slot and the $DIFS$ period values, respectively, and $q$ is the total number of times the channel was occupied by other stations during $T$.
Algorithm \ref{al:a2} we explains how to estimate backoff values for $S$. The algorithm can be configured to obtain backoff value selections for all stations in the WLAN. The output of the algorithm, i.e. vector $K$, will be used by CWE to estimate the \CW of $S$.
\begin{algorithm}[h]
  \caption{Backoff Estimation for $S$}
  \begin{algorithmic}[1]
  \STATE Input $T$ : Monitoring period ;\\
Variables: $q \leftarrow 0, idle \leftarrow 0, Q \leftarrow 1$;\\
Output: $K$;

  \WHILE{current time $< T$}
   \STATE Monitor the channel until it gets occupied
  \STATE $idle\leftarrow idle + \text{channel idle duration}$;
   \IF{ correct packet reception}
   	  \IF{ transmitter's MAC address match S's MAC address}
   	  	\IF{$Q$ has a nonezero value}
   	  	\STATE add $\frac{idle-q \times T_{DIFS}}{T_{MAC}}$ to $K$;
   	  	\STATE $idle,q\leftarrow 0$;
   	  	\ENDIF
   	  	\STATE $Q\leftarrow \text{\texttt{Queue Size} subfield value};$
 	 \ELSE
 	 	\STATE $q \leftarrow q+1$;
   	  \ENDIF
  \ELSE 
  	\STATE perform CIT to identify colliders
  	\IF{$S$ is a collider}
  		\STATE add $\frac{idle-q \times T_{DIFS}}{T_{MAC}}$ to $K$;
  		\STATE $idle,q\leftarrow 0$;
  	\ELSE
  	\STATE $q\leftarrow q+1$;
  	\ENDIF
   \ENDIF
  \ENDWHILE
  \end{algorithmic}
  \label{al:a2}
\end{algorithm}
\section{Collision Detection \& Identification Technique (CIT)}\label{sc:collision_detection}
In \cite{Gollakota08}, authors propose an algorithm to decode collided packets. For their algorithm to work, the number of distinct collisions (different overlapping combinations) that are needed to be gathered is the same as the number of colliding stations. Since, our backoff estimation algorithm only needs the ID's of colliding stations, we develop \textit{Collision Identification Technique} (CIT) in which a single collision is sufficient to identify all colliding stations without needing to decode packets. CIT uses wireless channel and hardware characteristics of each station-AP pair to identify stations involved in any collisions by following a correlation-based technique. Before going through details of CIT, we briefly present some background on digital communication and some physical phenomenons that affect signals transmitted over the wireless channel, namely, frequency offset, channel attenuation and channel phase shift.
\subsection{Digital Communications}
\begin{figure}
\includegraphics[scale=0.4]{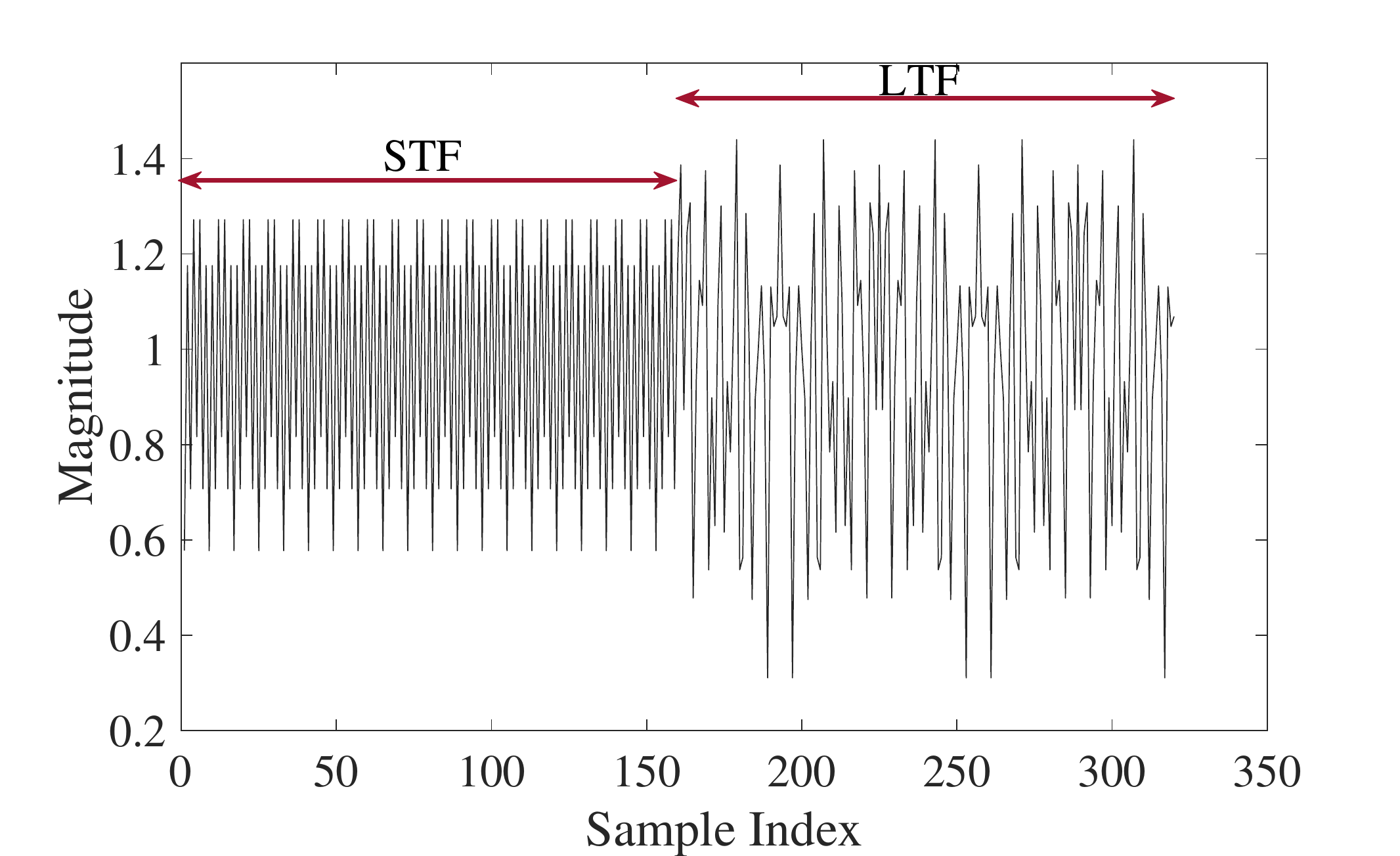}
\caption{Legacy preamble of an 802.11 packet.}
\label{fig:preamble}
\end{figure}

Packets are consisted of bits, for these bits to get transmitted over the wireless channel, they have to be modulated into complex stream of numbers. For example, the BPSK modulation scheme converts a bit of value $0$ and $1$ into complex symbols $e^{j\pi}=-1$ and $e^{j0}=1$, respectively. The transmitter generates symbols each $T_s$ seconds. We denote the $n$th symbol generated by the transmitter by $x(n)$. Considering there is only one transmitter, we denote the $n$th symbol received by a receiver with a sampling rate of $\frac{1}{T_s}$ by $y(n)$, which has the following relation with $x(n)$:
 \begin{equation}
y(n) = Hx(n)+\mathcal{N}(n),
    \label{eq:e1}
 \end{equation}
where $H=Me^{j\phi}$ is a complex number with a magnitude of $M$ and an angle of $\phi$, modeling  the channel attenuation and phase shift effects, respectively. Also, $\mathcal{N}$ models an AWGN channel. We consider the AP to be the receiver in our system model (uplink transmissions), which serves $N$ number of stations. Therefore, Equation \ref{eq:e1} can be generalized as follows:
 \begin{equation}
y(n) = \sum_{i=1}^{N}H_{i}x_{i}(n-\eta_i)u(n-\eta_i)+\mathcal{N}(n),
    \label{eq:e2}
 \end{equation}
where $H_i$ represents the channel between $S_i$ and the AP, and $x_{i}(n)$ represents the $n$th symbol transmitted by $S_i$, $u(n)$ is the unit step function, and $\eta_i$ is the index of the received symbol at the AP where $S_i$ starts transmitting.\\
For the AP to correctly receive transmitted symbols of $S_i$, it has to compensate for frequency offset (FO), sampling offset, inter-symbol interference, and channel equalization. However, for the purpose of collision detection we only need to explain the effects of FO, channel attenuation, and channel phase shift (PS) on transmitted symbols. \\
  FO is the absolute difference of transmitter and receiver oscillators' frequencies that are supposed to be centered as the exact same frequency.  The FO between a pair of transmitter-receiver results in a linear phase shift in received symbols that increases over time. The receiver usually estimates FO and compensates for it. As for PS, i.e. $e^{j\phi_i}$ in Equation \ref{eq:e2}, the phase of all symbols transmitted by $S_i$ is shifted by a value of $\phi_i$. The AP should compensate for the channel phase shift effect to correctly detect collisions. Typically, receivers estimate the channel response and compensate for the channel effects as they do for FO effects by using the 802.11 preamble \cite{80211}.  Equation \ref{eq:e2} can be further generalized to account for frequency offsets between AP and its stations as follows:
 \begin{equation}
y(n) = \sum_{i=1}^{N}H_{i}x_{i}(n-\eta_i)e^{j2\pi (n-\eta_i) \delta_{f}(i)T_{s}}u(n-\eta_i)+\mathcal{N}(n),
    \label{eq:e3}
 \end{equation}
 where $\delta_{f}(i)$ is the frequency offset between AP and $S_i$.
 CIT relies on studying the architecture of the 802.11 legacy preamble. Standardized preambles are designed to satisfy certain properties, including high FO estimation range, good frame detection accuracy, low dynamic range and low \textit{peak-to-average power ratio} (PAPR) \cite{hanif16}. Every PHY-layer frame starts with a preamble, which begins with two essential fields, short training field (STF) and long training field (LTF). Figure \ref{fig:preamble} shows the legacy preamble where the sampling frequency is 20 Msps. The STF contains ten identical short training sequences (STS's), which represent ten replicas of a particular periodic signal with period  $\lambda_{STF}= 0.8 \mu sec$. The STF is used for coerced FO estimation and frame detection \cite{Schmidl}. The LTF consists of two long training sequences (LTS's), which represent two cycles of another known periodic signal with period $\Delta_{LTF}= 4\Delta_{STF}$ , plus a 1.6 $\mu sec$ cyclic prefix. The LTF is used for channel estimation and further FO estimation. The legacy preamble is included in all the 802.11 enhancements. This is for the backward compatibility of the newer amendments with the legacy versions. For CIT to be applicable for all 802.11 versions, we will consider the legacy preamble to develop our algorithm, and refer to the legacy preamble as the ``preamble'', throughout the paper.

\subsection{Collision Identification Technique (CIT)}

\begin{figure}
\begin{center}
\includegraphics[scale=0.65]{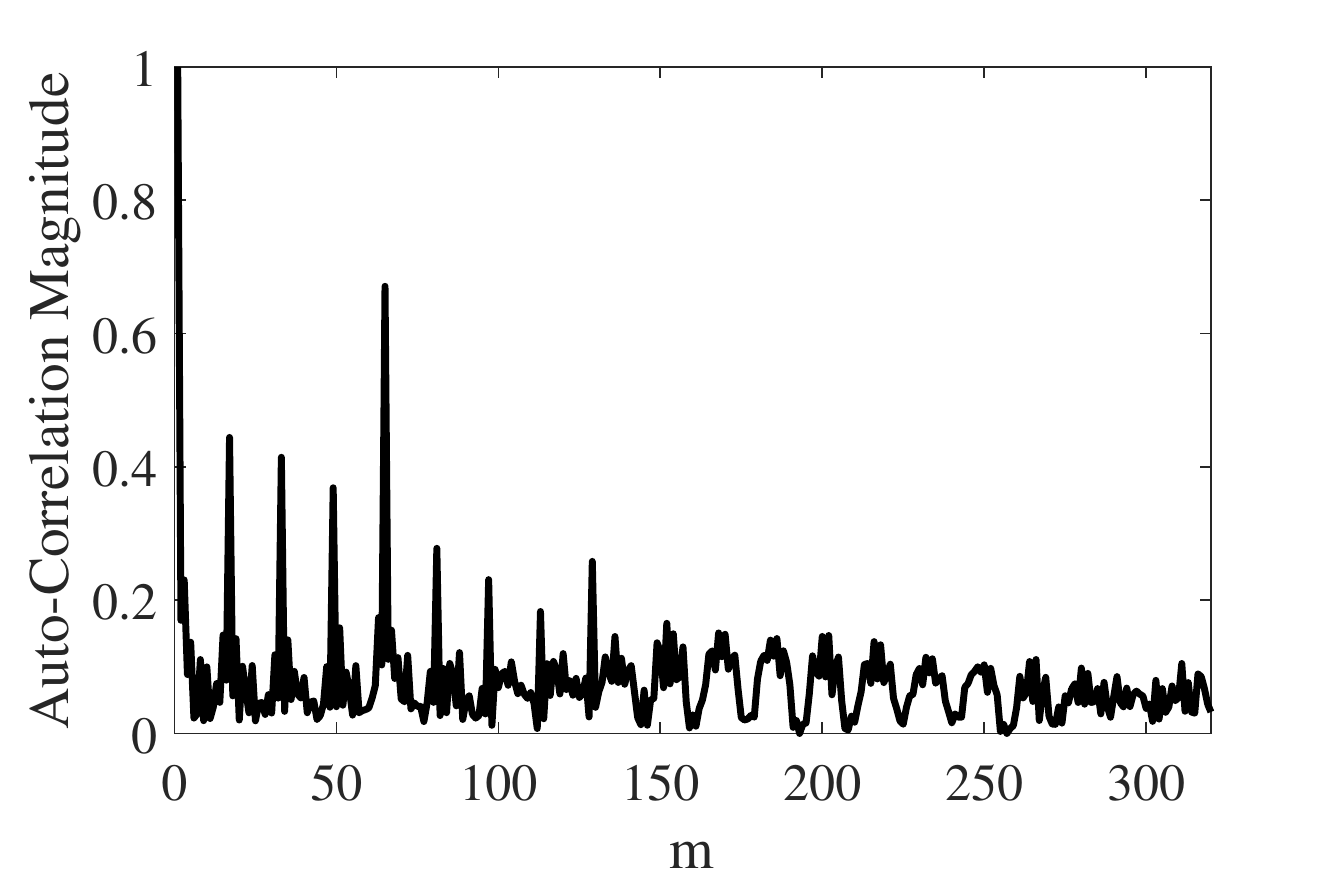}
\end{center}
\caption{Auto-correlation magnitude value of the 802.11 preamble.}
\label{fig:preauto}
\end{figure}

 If the AP receives a signal $y$ that it fails to correctly decode, it will initiate CIT. The heart of our collision detection technique is to leverage the fact that the 802.11 packets start with a known set of samples (i.e., 802.11 preamble). CIT uses this fact and calculates the cross-correlation of the known preamble, which is modified to incorporate the FO and channel PS effects, with the collided signal, and looks for peaks in this cross-correlation that exceed a detection threshold. Equation \ref{eq:e3} shows that FO and PS effects change the phase of the transmitted symbols, hence the cross-correlation might not peak where the known preamble overlaps with the start of a Wi-Fi transmission. We overcome the former challenge by requiring the AP to keep the latest FO and PS estimations of each successful transmission in the vectors $\delta_f=[\delta_{f}(1),\delta_{f}(2),..., \delta_f(N)]$ and  $\phi =[\phi(1),...,\phi(N)]$, respectively, where $\delta_f(i)$ and $\phi(i)$ are the latest estimated FO and PS between station $S_i$ and the AP, respectively. Upon receiving a collided signal $y$,  the AP first, modifies the known preamble by incorporating the effects of FO and PS on the known preamble, then it computes the cross-correlation with $y$. The modified preamble is denoted by $P_i$ and obtained as:

\begin{equation}
P_i(n)=e^{j\phi (i)}e^{j2\pi n \delta_f(i)T_s}P(n)\quad n=1,..,L,
\label{eq:pre}
\end{equation}
where $P$ is the original preamble and $L$ is its length. The cross-correlation between $P_i$ and $y$ is denoted by $\Gamma_i$ and can be calculated as:
\begin{itemize}
\item[(i)] $0 \leq m \leq |y|$:
 \begin{equation}
 \begin{aligned}
& \Gamma_i(m)=\\
&\frac{\left| \sum_{n=1}^{L}P_{i}^{*}(n) y(n+m)\right |}{\sqrt{ \sum_{n=1}^{L}P_{i}^{*}(n) P_{i}(n) }\sqrt{ \sum_{n=1}^{L}y^{*}(n+m)  y(n+m) }} 
\end{aligned}
    \label{eq:e4_1}
 \end{equation}
 
\item[(ii)] $ m <0 \quad \text{or} \quad m > |y|$:
 \begin{equation}
\Gamma_i(m) =0,
    \label{eq:e4_2}
 \end{equation}
\end{itemize}

where ``$*$'' is the complex conjugate operator. Also we zero-pad $y$ to gather cross-correlation results for $|y|-L<m<|y|$. 
 In Equation \ref{eq:e4_1}, the two factors in the denominator are normalizing the value of $\Gamma_i$ for it to be in the range $[0,1]$. To keep track of the highest cross-correlation values over all $\Gamma_i$'s for each received symbol, $m \in \{0,1,...,|y|\}$), the AP builds a composite cross-correlation vector $\Gamma$ as:
 \begin{equation}
\Gamma (m) =max(\Gamma_1(m),...,\Gamma_N(m)), m \in \{0,1,...,|y|\}.
    \label{eq:gamma}
 \end{equation}
 Then, each $m$ that has a $\Gamma(m)>th_c$ will be assigned to a station to be identified as a collider. In our analysis of 802.11 packets we have seen that if a station, say $S_i$, is a collider then $\Gamma_i$ will have comparable cross-correlation values for values of $m$ that are surrounding $\eta_i$, i.e., the index of the first received symbol transmitted by $S_i$. This is due to preamble's periodic nature. In Figure \ref{fig:preauto}, we show the auto-correlation of the 802.11 preamble while having a sampling rate of 20 Msps. We can see that several high-value peaks reside in the neighborhood of the start of the preamble. And from Equation \ref{eq:gamma}, we can see that only the highest values of cross-correlation will be considered for each index of $\Gamma$. So, to prevent the false detection of stations as colliders and falsely not detecting colliding stations, we need to eliminate the surrounding cross-correlation peak values around  highest peak values. We realize the former by defining a filtering window of size $\zeta$ that assigns a value of zero for cross-correlation values of $m$'s that are within the $\zeta$ range of the highest cross-correlation values. CIT's filtering approach proceeds dynamically by first zeroing out the cross-correlation values of indices surrounding the index of highest cross-correlation value, then updating $\Gamma_j$'s, and doing the former for the second highest peak, and proceeding similarly for all the remaining cross-correlation values higher than $th_c$. We are assuming that in a collision, packets sent by different stations are apart from one another at least by a MAC time slot $T_{MAC}$. Therefore, $\zeta=\frac{T_{MAC}}{T_s}$ is a reasonable choice.  After filtering, the AP constructs the composite  cross-correlation vector $\Gamma$ by the updated $\Gamma_i$'s using Equation \ref{eq:gamma}, once more. Then each value in $\Gamma$ that is higher than a detection threshold $th_c$ is assigned to the station with that specific cross-corelation value at that specific received symbol index. Each station with an assigned peak value will be considered as a participant transmitter for the received signal $y$. After assigning all the peak values, CIT will construct a vector, $ID$, that will consist of the IDs of all the colliding stations during $y$.
%
%
%

\begin{figure*}[h]
\centering
\includegraphics[scale=0.22]{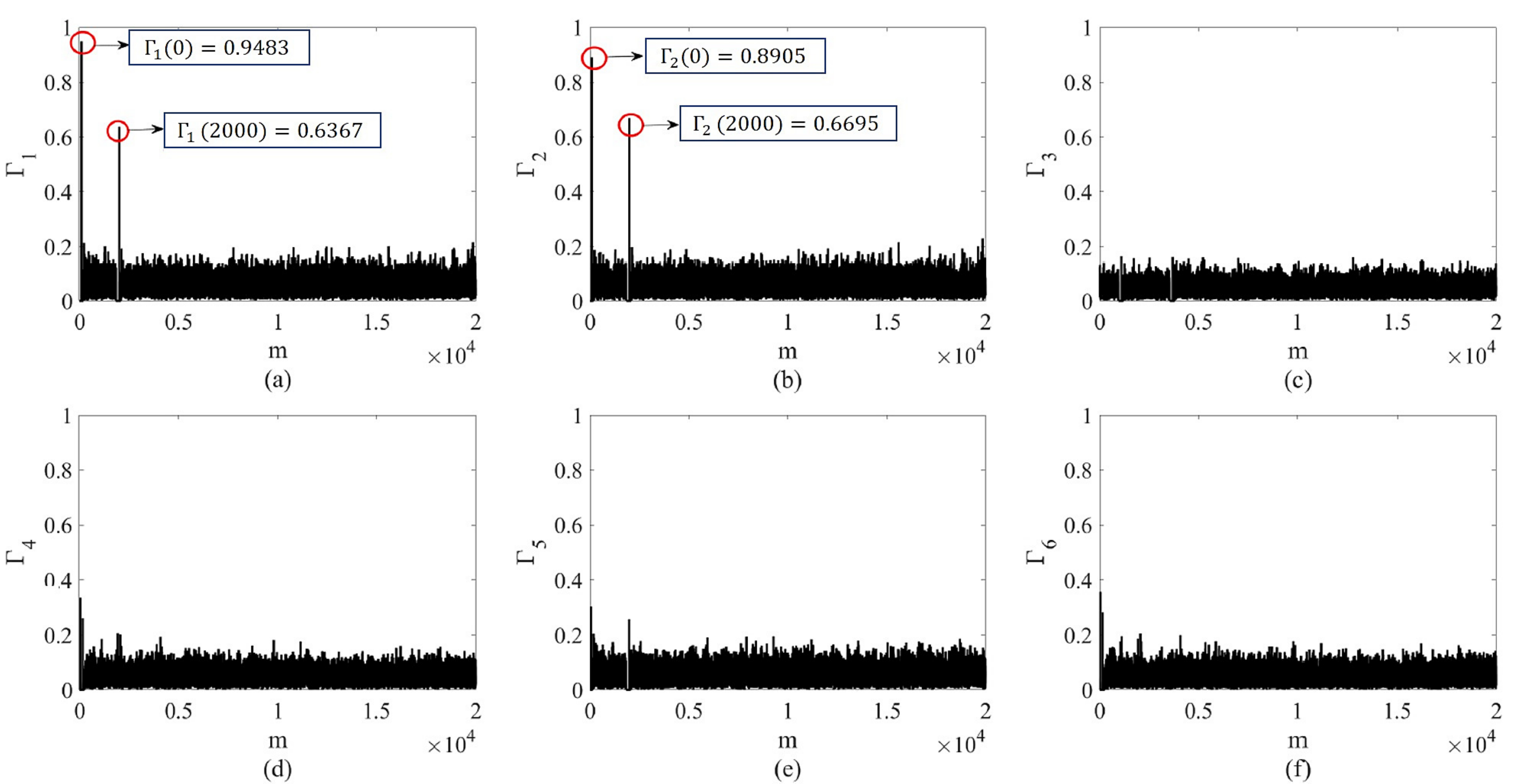}
\caption{The cross-correlation of the modified preambles with the collided signal, $y$ for (a)$S_1$, (b)$S_2$, (c) $S_3$, (d) $S_4$, (e) $S_5$, and (f) $S_6$.}
\label{fig:corr}
\end{figure*}

To better understand the process of CIT, we provide a simulation example using MATLAB's Wireless Waveform Generator \cite{Matlab}. Consider a WLAN with six stations and one AP, where $S_1$ and $S_2$ are hidden terminals. $S_1$ starts its transmission and the AP senses the channel to busy and start receiving the transmitting signal, during $S_1$'s transmission, all stations except $S_2$ freeze their backoff counter. After the AP receives about 2000 samples, with a sampling rate of 20 Msps, $S_2$ starts its transmission, while the AP continues receiving samples but it will not be able to successfully decode any packets. So it initiates CIT which starts by calculating the cross-correlation values for each station and adopting a filtering of $\zeta=\frac{T_{MAC}}{T_s}=\frac{9 \mu s}{\frac{1}{20 Msps}}=180$. In Figure \ref{fig:corr}, we present the magnitude of the cross-correlation values of the modified preambles with $y$ for 6 stations, where each station is 5 meters away from the AP. We use itu-r m.2135-1 channel path loss model with $SNR=10$ dB. The center frequency and bandwidth for both transmission and reception are 2.4 GHz and 20 MHz, respectively. We randomly select the elements of the vectors $\delta_f$ and $\phi$ from the intervals $[-125 ,125 ]$ KHz, and $[0,2\pi]$, respectively. We set $\zeta=180$ and $th_c=0.6$. It can be seen that all $\Gamma_i$'s have peak values at $m$ indices that correspond to the start of a Wi-Fi transmission. However, the highest cross-correlation value is for the $\Gamma_i$ that is correctly modifying the transmitted preamble (incorporating the right values for FS and PS in Equation \ref{eq:pre}). Looking at Figure \ref{fig:corr}, $\Gamma_1$ has the largest cross-correlation value at $m=0$ ($\Gamma_1(0)=0.9483$), hence $\Gamma(0)=\Gamma_1(0)$. Following the same procedure and constructing $\Gamma$ for the remaining $m$ values, it can be seen that $\Gamma$ will have only one other value larger than $th_c$, which is at $m=2000$ ($\Gamma(2000)=0.6695$). $\Gamma(2000)$ is associated to $S_2$, since $\Gamma_2(2000)=\Gamma(2000)>th_c$. Since there are no other $\Gamma(m)$'s larger than $th_c$, CIT will terminate with $ID=\{S_1,S_2\}$, with $S_1$ and $S_2$ having transmission start indices of $m=0$ and $m=2000$ during $y$, respectively. .\\

\section{Evaluations} \label{sc:evalualtions}

\subsection{Colision Identification technique}
To obtain the accuracy of CIT, we conduct simulations, using MATLAB's  WLAN toolbox, for three settings of a WLAN with $N=3$, $6$, and $9$. In all the settings, stations $S_1$ and $S_2$ are hidden terminals. We set the center frequency and bandwidth for transmission of all stations to 2.4 GHz and 20 MHz, respectively. Our algorithm's performance is dependent on $th_c$ and $\zeta$. Therefore, we vary $th_c$ from 0 to 1 and set $\zeta=180$. For each $th_c$ value, we run 100 different simulations, with different random seeds, each including 1000 different collision combinations of $S_1$ and $S_2$. We derive the accuracy as:
 \begin{equation}
\text{Accuracy}=\frac{\text{Number of correct detections}}{\text{Total number of detections}}\times 100 \%,
    \label{eq:ea}
 \end{equation}
 where a correct detection translates into correct identification of all colliders in $y$. We randomly select the PS values of all stations to be in the range $[0,2\pi]$. Also the FO values of all stations is randomly selected from $[-125,125]$ kHz for each simulation run, which is the acceptable FO for 2.4 GHz center frequency \cite{80211}.  It is important to note that as the FO values of stations get closer to each other the possibility of a miss-detection increases. To fully illustrate the effect of the former, we include a new parameter $\Delta$ into our evaluations, which effects the random selection of FO values. The value of $\Delta$ indicates that for any element of $\delta_f$, e.g. $\delta_f(i)$, the only element of $\delta_f$ residing in the frequency range $[\delta_f(i)-\frac{\Delta}{100}\delta_f(i),\delta_f(i)+\frac{\Delta}{100}\delta_f(i)]$ is $\delta_f(i)$.

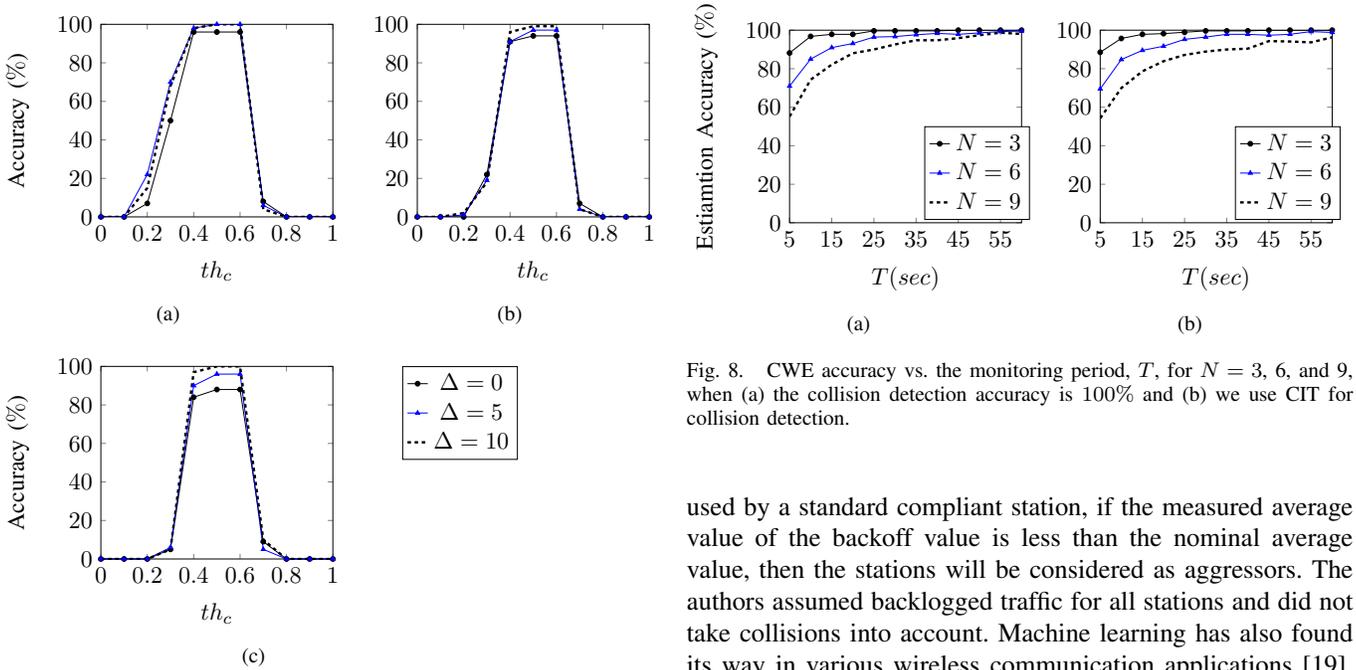
\begin{figure}[h]
\subfigure[]{
\begin{tikzpicture}[scale=0.45, transform shape]
\begin{axis}[title={},xlabel={$th_c$},
    ylabel={Accuracy ($\%$)},
    xmin=0, xmax=1,
    ymin=0, ymax=100,
   nodes={scale=2, transform shape}]

\addplot[color=black,mark=*]coordinates{(0,0)(0.1,0)(0.2,7)(0.3,50)(0.4,96)(0.5,96)(0.6,96)(0.7,8)(0.8,0)(0.9,0)(1,0)
};

\addplot[color=blue,mark=triangle*]coordinates{(0,0)(0.1,0)(0.2,22)(0.3,70)(0.4,98)(0.5,100)(0.6,100)(0.7,6)(0.8,0)(0.9,0)(1,0)
};

 \addplot [color=black, dashed,line width=2pt] coordinates {(0,0)(0.1,0)(0.2,15)(0.3,68)(0.4,98)(0.5,100)(0.6,100)(0.7,4)(0.8,0)(0.9,0)(1,0)};
 
\end{axis}
\end{tikzpicture}
  \label{fig:vt3}

}\subfigure[]{
\begin{tikzpicture}[scale=0.45, transform shape]
\begin{axis}[title={},xlabel={$th_c$},
    xmin=0, xmax=1,
    ymin=0, ymax=100,
    nodes={scale=2, transform shape}]

\addplot[color=black,mark=*]coordinates{(0,0)(0.1,0)(0.2,0)(0.3,22)(0.4,91)(0.5,94)(0.6,94)(0.7,7)(0.8,0)(0.9,0)(1,0)
};

\addplot[color=blue,mark=triangle*]coordinates{(0,0)(0.1,0)(0.2,1)(0.3,19)(0.4,91)(0.5,97)(0.6,97)(0.7,4)(0.8,0)(0.9,0)(1,0)
};

 \addplot [color=black, dashed,line width=2pt] coordinates {(0,0)(0.1,0)(0.2,2)(0.3,18)(0.4,96)(0.5,99)(0.6,99)(0.7,4)(0.8,0)(0.9,0)(1,0)};
 
\end{axis}
\end{tikzpicture}
  \label{fig:vt6}

}\\

\subfigure[]{
\begin{tikzpicture}[scale=0.45, transform shape]
\begin{axis}[title={},xlabel={$th_c$},
    ylabel={Accuracy ($\%$)},
    xmin=0, xmax=1,
    ymin=0, ymax=100,
    legend style={at={(axis cs:1.3,100)},anchor=north west}
    ,nodes={scale=2, transform shape}]

\addplot[color=black,mark=*]coordinates{(0,0)(0.1,0)(0.2,0)(0.3,5)(0.4,84)(0.5,88)(0.6,88)(0.7,9)(0.8,0)(0.9,0)(1,0)
};
\addlegendentry{$\Delta=0$}

\addplot[color=blue,mark=triangle*]coordinates{(0,0)(0.1,0)(0.2,0)(0.3,6)(0.4,90)(0.5,96)(0.6,96)(0.7,5)(0.8,0)(0.9,0)(1,0)
};
\addlegendentry{$\Delta=5$}

 \addplot [color=black, dashed,line width=2pt] coordinates {(0,0)(0.1,0)(0.2,0)(0.3,5)(0.4,97)(0.5,100)(0.6,100)(0.7,10)(0.8,0)(0.9,0)(1,0)};
 \addlegendentry{$\Delta=10$};
 
\end{axis}
\end{tikzpicture}
  \label{fig:vt9}

}
  \caption{Accuracy of CIT  vs. the collision detection threshold (i.e., $th_c$) for (a) $N=3$, (b) $N=6$, and (c) $N=9$. }
  \label{fig:vt}
\end{figure}
In Figures \ref{fig:vt3}, \ref{fig:vt6}, and \ref{fig:vt9}, we show CIT's accuracy vs. $th_c$ for $N=3$, $6$, and $9$, respectively. It can be seen that for $th_c=0.5$, CIT can achieve $96\%$, $94\%$, $88\%$ collision identification accuracy for when we have $N=3$, 6, and 9, respectively.

\subsection{\CW Estimation}
Our simulation evaluations are based on a C++-based discrete-event simulator called CSIM \cite{CSIM_lib}. CSIM includes functions and classes for generating and synchronizing process-oriented events. We implement the DCF as detailed in 802.11 ac standard, including all timing requirements. An indoor scenario is considered, where a number of Wi-Fi devices are uniformly distributed in a square area of length 80 meters. In this section we  present \CW estimation (CWE) accuracy results for WLAN's where $N=3$, $6$, and $9$. In all our simulation settings, the \CW value of all stations are randomly selected from $\{2,3,...,16\}$. We conduct 93, 70, and 51 simulation setups for $N=3$, 6, and 9, respectively, which results into 279, 420, and 459 total estimations in total. In Figures \ref{fig:estac} and \ref{fig:estac_reality} , we show the accuracy performance of CWE vs. the monitoring period (i.e., $T$) for when we have a collision detection accuracy of $100 \%$ and for when we use CIT, respectively. It can be seen that by using CIT with  $T=60 $ sec we  can achieve $100\%$, $98.81\%$, and $96.3 \%$ CWE accuracy, for when we have $N=3$, 6, and 9, respectively.

\begin{figure}[h]
\subfigure[]{\begin{tikzpicture}[scale=0.45, transform shape]
\begin{axis}[title={},xlabel={$T (sec)$},
    ylabel={Estiamtion Accuracy ($\%$)},
    xmin=5, xmax=60,
    ymin=0, ymax=100,
    xtick={5,15,25,35,45,55},
    xtick={5,15,25,35,45,55},
    legend style={at={(axis cs:37,50)},anchor=north west}
    ,nodes={scale=2, transform shape}]

\addplot[color=black,mark=*]coordinates{(5,88.17204301075269)(10,96.7741935483871)(15,97.84946236559139)(20,97.84946236559139)(25,99.6415770609319)(30,99.6415770609319)(35,99.6415770609319)(40,99.6415770609319)(45,100.0)(50,100.0)(55,100.0)(60,100.0)
};
\addlegendentry{$N=3$}

\addplot[color=blue,mark=triangle*]coordinates{(5,70.95238095238095)(10,85.0)(15,90.95238095238095)(20,93.0952380952381)(25,96.42857142857143)(30,96.66666666666667)(35,97.61904761904762)(40,98.33333333333333)(45,97.85714285714285)(50,98.57142857142858)(55,99.28571428571429)(60,99.52380952380952)
};
\addlegendentry{$N=6$}

 \addplot [color=black, dashed,line width=2pt] coordinates {(5,55.119825708061)(10,74.29193899782135)(15,82.13507625272331)(20,88.01742919389977)(25,89.97821350762527)(30,92.5925925925926)(35,94.77124183006535)(40,94.77124183006535)(45,95.86056644880175)(50,97.38562091503267)(55,98.69281045751634)(60,98.0392156862745)};
 \addlegendentry{$N=9$};
 
\end{axis}
\end{tikzpicture}
  \label{fig:estac}
}\subfigure[]{
\begin{tikzpicture}[scale=0.45, transform shape]
\begin{axis}[title={},xlabel={$T (sec)$},
    ylabel={},
    xmin=5, xmax=60,
    ymin=0, ymax=100,
    xtick={5,15,25,35,45,55},
    xtick={5,15,25,35,45,55},
    legend style={at={(axis cs:37,50)},anchor=north west}
    ,nodes={scale=2, transform shape}]

\addplot[color=black,mark=*]coordinates{(5,88.5304659498208)(10,95.6989247311828)(15,97.84946236559139)(20,98.2078853046595)(25,98.9247311827957)(30,99.6415770609319)(35,99.6415770609319)(40,99.6415770609319)(45,100.0)(50,100.0)(55,100.0)(60,100.0)
};
\addlegendentry{$N=3$}

\addplot[color=blue,mark=triangle*]coordinates{(5,69.52380952380952)(10,84.76190476190476)(15,89.52380952380953)(20,91.66666666666666)(25,95.23809523809523)(30,96.42857142857143)(35,97.85714285714285)(40,97.85714285714285)(45,97.38095238095238)(50,97.85714285714285)(55,99.28571428571429)(60,98.80952380952381)
};
\addlegendentry{$N=6$}

 \addplot [color=black, dashed,line width=2pt] coordinates {(5,54.248366013071895)(10,69.93464052287581)(15,78.64923747276688)(20,83.87799564270153)(25,87.14596949891067)(30,88.88888888888889)(35,89.97821350762527)(40,90.41394335511983)(45,94.33551198257081)(50,94.11764705882352)(55,93.68191721132898)(60,96.29629629629629)
};
 \addlegendentry{$N=9$};

\end{axis}
\end{tikzpicture}
  \label{fig:estac_reality}
}
  \caption{ CWE accuracy  vs. the monitoring period, $T$, for $N=3$, 6, and 9, when (a) the collision detection accuracy is $100\%$ and (b) we use CIT for collision detection.}
  \end{figure}

\section{Related Works} \label{sc:relatedworks}
In \cite{Rong06}, Rong et al.'s misbehavior detection scheme is based on the sequential hypothesis testing. Instead of monitoring the  backoff values selected by stations they first developed analytical models for packet inter-arrival time distribution from each station in the network, where multiple cheating stations coexist. Using the characteristics of this probability distribution, they developed an algorithm to detect cheating stations based on the throughput degradations observed at normal stations. However, they only considered saturated traffic and they assumed that all stations are implementing RTS/CTS exchange. To detect misbehavior in 802.11 WLAN's, Tang et al. \cite{Tang14} assumed that the number of aggressors in the WLAN is known and they derived Markov chains for different settings of aggressors and well-behaved stations, then they analyzed the successful transmission rate of the tagged station to see whether it will reach beyond the rate of a standard station's to be considered as an aggressor. The authors assumed that stations are saturated with traffic. Also, they only assumed one aggressor and considered stations to be in each others sensing ranges, hence eliminating the chance of collisions and hidden terminals. The authors in \cite{Li15}, proposed mechanisms to detect and penalize aggressors that choose a low \cw . The detection is applied on multiple observations of backoff values and then compared to a supposed average backoff value to determine whether it is less than the supposed value, if so, then the station is considered as an aggressor. Each station is observed by all its one hop neighbors and all stations are considered to have backlogged traffic. The authors claimed that the hidden terminal problem is solved by taking the majority vote for deciding whether a station is aggressive. For the AP to detect stations with low \CW values, Raya et al. \cite{Raya06} proposed that the AP should first gather backoff value traces from each station and then compare the average value of selected backoff values with a nominal average backoff value which is used by a standard compliant station, if the measured average value of the backoff value is less than the nominal average value, then the stations will be considered as aggressors. The authors assumed backlogged traffic for all stations and did not take collisions into account. Machine learning has also found its way in various wireless communication applications \cite{AmirMohammed2019DySPAN, amircnn,amirinfo}. In \cite{AmirMohammed2019DySPAN}, the authors tackle the aggressive behavior of stations in the WLAN by equipping the standard stations with a machine learning module, specifically random forests, to adapt their \CW to get their fair share of channel airtime. Their framework only enhances the performance of stations that utilize their module, thus the performance of standard-compliant stations that do not use their adaptation algorithm might decrease. Also, for collision detection, the work closest to ours is \cite{Gollakota08}, which we introduced in section
\ref{sc:collision_detection}. Zhao et al \cite{Zhao20} have also developed an algorithm to resolve RTS collisions, by analyzing the payload of the RTS as a vector and obtaining its distribution,  and reformulating the RTS resolution as a sparse-recovery problem. 
\section{Conclusions} \label{sc:conclusions}
In this work, we showed the unfairness that will be created when aggressive stations with low \CW exist in the WLAN. We proposed a novel solution for the AP to detect aggressors in the WLAN by estimating their \cw 's, i.e., CWE. Using CWE, the AP needed  to monitor the backoff values of its stations for \CW estimation, which required the AP to keep track of the idle time each station spent backing off. The former also needed the AP to be able identify colliding stations in an 802.11 uplink collision, which we tried to resolve by introducing our collision detection and identification technique (CIT).  Overall, our collision detection algorithm obtains accuracies of $96 \%$, $94\%$, and $88\%$, our \cw -estimation algorithm has estimation accuracies of $100\%$, $98.81\%$, and $96.3\%$,  when we have 3, 6, and 9 stations in the WLAN, respectively. 

\section{Acknowledgments} \label{sc:conclusions}
This research was supported in part by NSF (grants CNS-1563655, CNS-1731164, and IIP-1822071), and in part by the U.S. Army Small Business Innovation Research Program Office and Army Research Office under Contract No. W911NF-21-C-0016. Any opinions, findings, conclusions, or recommendations expressed in this paper are those of the author(s) and do not necessarily reflect the views of NSF or Army.

\bibliographystyle{IEEEtran}

\bibliography{The_bilbiography}

\begin{thebibliography}{10}
\providecommand{\url}[1]{#1}
\csname url@samestyle\endcsname
\providecommand{\newblock}{\relax}
\providecommand{\bibinfo}[2]{#2}
\providecommand{\BIBentrySTDinterwordspacing}{\spaceskip=0pt\relax}
\providecommand{\BIBentryALTinterwordstretchfactor}{4}
\providecommand{\BIBentryALTinterwordspacing}{\spaceskip=\fontdimen2\font plus
\BIBentryALTinterwordstretchfactor\fontdimen3\font minus
  \fontdimen4\font\relax}
\providecommand{\BIBforeignlanguage}[2]{{%
\expandafter\ifx\csname l@#1\endcsname\relax
\typeout{** WARNING: IEEEtran.bst: No hyphenation pattern has been}%
\typeout{** loaded for the language `#1'. Using the pattern for}%
\typeout{** the default language instead.}%
\else
\language=\csname l@#1\endcsname
\fi
#2}}
\providecommand{\BIBdecl}{\relax}
\BIBdecl

\bibitem{80211}
``{IEEE} standard for information technology—telecommunications and
  information exchange between systems local and metropolitan area
  networks—specific requirements - part 11: Wireless lan medium access
  control (mac) and physical layer (phy) specifications,'' \emph{IEEE Std
  802.11-2016 (Revision of IEEE Std 802.11-2012)}, pp. 1--3534.

\bibitem{Xu05}
W.~Xu, W.~Trappe, Y.~Zhang, and T.~Wood, ``The feasibility of launching and
  detecting jamming attacks in wireless networks,'' in \emph{In Proc. of the
  ACM MobiHoc}, 2005, p. 46–57.

\bibitem{toledo08}
A.~L. {Toledo} and X.~{Wang}, ``Robust detection of mac layer denial-of-service
  attacks in csma/ca wireless networks,'' \emph{IEEE Transactions on
  Information Forensics and Security}, vol.~3, no.~3, pp. 347--358, 2008.

\bibitem{manzo05}
M.~Manzo, T.~Roosta, and S.~Sastry, ``Time synchronization attacks in sensor
  networks,'' in \emph{Proceedings of the 3rd ACM Workshop on Security of Ad
  Hoc and Sensor Networks}, 2005, p. 107–116.

\bibitem{raymond09}
D.~R. {Raymond}, R.~C. {Marchany}, M.~I. {Brownfield}, and S.~F. {Midkiff},
  ``Effects of denial-of-sleep attacks on wireless sensor network mac
  protocols,'' \emph{IEEE Transactions on Vehicular Technology}, vol.~58,
  no.~1, pp. 367--380, 2009.

\bibitem{Li15}
M.~{Li}, S.~{Salinas}, P.~{Li}, J.~{Sun}, and X.~{Huang}, ``Mac-layer selfish
  misbehavior in ieee 802.11 ad hoc networks: Detection and defense,''
  \emph{IEEE Transactions on Mobile Computing}, vol.~14, no.~6, pp. 1203--1217,
  2015.

\bibitem{radosavac05}
S.~Radosavac, J.~S. Baras, and I.~Koutsopoulos, ``A framework for mac protocol
  misbehavior detection in wireless networks,'' in \emph{Proceedings of the 4th
  ACM Workshop on Wireless Security}, 2005, p. 33–42.

\bibitem{Kyasanur05}
P.~{Kyasanur} and N.~H. {Vaidya}, ``Selfish mac layer misbehavior in wireless
  networks,'' \emph{IEEE Transactions on Mobile Computing}, vol.~4, no.~5, pp.
  502--516, 2005.

\bibitem{lu10}
Z.~{Lu}, W.~{Wang}, and C.~{Wang}, ``On order gain of backoff misbehaving nodes
  in csma/ca-based wireless networks,'' in \emph{2010 Proceedings IEEE
  INFOCOM}, 2010, pp. 1--9.

\bibitem{law09}
Y.~W. Law, M.~Palaniswami, L.~V. Hoesel, J.~Doumen, P.~Hartel, and P.~Havinga,
  ``Energy-efficient link-layer jamming attacks against wireless sensor network
  mac protocols,'' \emph{ACM Trans. Sen. Netw.}, 2009.

\bibitem{khatib10}
K.~{El-Khatib}, ``Impact of feature reduction on the efficiency of wireless
  intrusion detection systems,'' \emph{IEEE Transactions on Parallel and
  Distributed Systems}, pp. 1143--1149, 2010.

\bibitem{gupta02}
V.~{Gupta}, S.~{Krishnamurthy}, and M.~{Faloutsos}, ``Denial of service attacks
  at the mac layer in wireless ad hoc networks,'' in \emph{In Proc. of the IEEE
  MILCOM}, 2002, pp. 1118--1123.

\bibitem{konorski01}
J.~Konorski, ``Protection of fairness for multimedia traffic streams in a
  non-cooperative wireless lan setting,'' in \emph{In Proc. of the PROMS},
  2001, pp. 116--129.

\bibitem{konorski02}
------, ``Multiple access in ad-hoc wireless lans with noncooperative
  stations,'' in \emph{In Proc. of the NETWORKING}, 2002, pp. 1141--1146.

\bibitem{cagalj04}
M.~Cagalj, S.~Ganeriwal, I.~Aad, and J.-P. Hubaux, ``On cheating in csma/ca ad
  hoc networks,'' Tech. Rep., 2004.

\bibitem{Rong06}
Y.~{Rong}, S.~. {Lee}, and H.~. {Choi}, ``Detecting stations cheating on
  backoff rules in 802.11 networks using sequential analysis,'' in \emph{Proc.
  of the IEEE INFOCOM}, 2006, pp. 1--13.

\bibitem{Tang14}
J.~{Tang}, Y.~{Cheng}, and W.~{Zhuang}, ``Real-time misbehavior detection in
  ieee 802.11-based wireless networks: An analytical approach,'' \emph{IEEE
  Transactions on Mobile Computing}, vol.~13, no.~1, pp. 146--158, 2014.

\bibitem{Raya06}
M.~{Raya}, I.~{Aad}, J.~. {Hubaux}, and A.~{El Fawal}, ``Domino: Detecting mac
  layer greedy behavior in ieee 802.11 hotspots,'' \emph{IEEE Transactions on
  Mobile Computing}, vol.~5, no.~12, pp. 1691--1705, 2006.

\bibitem{AmirMohammed2019DySPAN}
A.~H.~Y. {Abyaneh}, M.~{Hirzallah}, and M.~{Krunz}, ``Intelligent-{CW}:
  {AI}-based {F}ramework for {C}ontrolling {C}ontention {W}indow in {WLAN}s,''
  in \emph{Proc. of the IEEE DySPAN}, 2019, pp. 1--10.

\bibitem{bianchi00}
G.~{Bianchi}, ``Performance analysis of the ieee 802.11 distributed
  coordination function,'' \emph{IEEE Journal on Selected Areas in
  Communications}, vol.~18, no.~3, 2000.

\bibitem{cha08}
S.-H. Cha, ``Taxonomy of nominal type histogram distance measures,'' in
  \emph{In Proc. of the American Conference on Applied Mathematics}, 2008, p.
  325–330.

\bibitem{Gollakota08}
S.~Gollakota and D.~Katabi, ``Zigzag decoding: Combating hidden terminals in
  wireless networks,'' in \emph{Proc. of the ACM SIGCOMM 2008 Conference on
  Data Communication}, 2008, p. 159–170.

\bibitem{hanif16}
H.~{Rahbari} and M.~{Krunz}, ``Rolling preambles: Mitigating stealthy fo
  estimation attacks in ofdm-based 802.11 systems,'' in \emph{In Proc. of the
  {IEEE} CNS}, 2016, pp. 118--126.

\bibitem{Schmidl}
T.~M. {Schmidl} and D.~C. {Cox}, ``Robust frequency and timing synchronization
  for ofdm,'' \emph{{IEEE} Transactions on Communications}, pp. 1613--1621,
  1997.

\bibitem{Matlab}
``{WLAN} toolbox version 3.0,'' R2020a, the MathWorks, Natick, MA, USA.

\bibitem{CSIM_lib}
``Csim20,'' [http://www.mesquite.com], accessed: 2019-08-30.

\bibitem{amircnn}
A.~Yazdani~Abyaneh, V.~Pourahmadi, and A.~Hosein Gharari~Foumani, ``Csi-based
  authentication: Extracting stable features using deep neural networks,''
  \emph{Transactions on Emerging Telecommunications Technologies}, vol.~31,
  no.~2, 2020.

\bibitem{amirinfo}
W.~Zhang, M.~Feng, M.~Krunz, and A.~H. Yazdani~Abyaneh, ``Signal detection and
  classification in shared spectrum: A deep learning approach,'' in \emph{Proc.
  of the IEEE INFOCOM}, 2021, pp. 1--10.

\bibitem{Zhao20}
S.~Zhao, Z.~Qu, Z.~Luo, Z.~Lu, and Y.~Liu, ``Comb decoding towards
  collision-free {WiFi},'' in \emph{Proc. of the {NSDI}}, 2020, pp. 933--951.

\end{thebibliography}

\end{document}